# Experimental Evidence of Near-field Superluminally Propagating Electromagnetic Fields

William D. Walker
Royal Institute of Technology, KTH-Visby
Department of Electrical Engineering
Cramérgatan 3, S-621 57 Visby, Sweden
bill@visby.kth.se

## 1 Introduction

A simple experiment is presented which indicates that electromagnetic fields propagate superluminally in the near-field next to an oscillating electric dipole source. A high frequency 437MHz, 2 watt sinusoidal electrical signal is transmitted from a dipole antenna to a parallel near-field dipole detecting antenna. The phase difference between the two antenna signals is monitored with an oscilloscope as the distance between the antennas is increased. Analysis of the phase vs distance curve indicates that superluminal transverse electric field waves (phase and group) are generated approximately one-quarter wavelength outside the source and propagate toward and away from the source. Upon creation, the transverse waves travel with infinite speed. The outgoing transverse waves reduce to the speed of light after they propagate about one wavelength away from the source. The inward propagating transverse fields rapidly reduce to the speed of light and then rapidly increase to infinite speed as they travel into the source. The results are shown to be consistent with standard electrodynamic theory.

Theoretical analysis of an oscillating electric dipole reveals that the longitudinal component of the electric field and the transverse magnetic field are generated at the source and propagate away from the source. Upon creation, the waves travel with infinite speed and then rapidly reduce to the speed of light after they propagate about one wavelength away from the source. It is noted that the special theory of relativity predicts that from a moving reference frame superluminal signals can propagate backward in time. Arguments against the superluminal wave interpretation presented in this paper are reviewed and shown to be invalid. Because of the similarity of the governing partial differential equations, two other physical systems (magnetic dipole and a gravitationally radiating oscillating mass) are noted to have similar superluminal near-field theoretical results.

## 2 Theoretical expectations from electromagnetic theory

### 2.1 Electromagnetic theoretical solution of oscillating electric dipole

Numerous textbooks present solutions of the electromagnetic (EM) fields generated by an oscillating electric dipole. The resultant electrical and magnetic field components for an oscillating electric dipole are known to be [1, 2]:

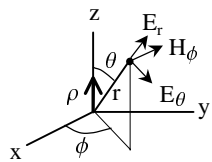

$$E_r = \frac{\rho Cos(\theta)}{2\pi\varepsilon_o \, r^3}[1 - i(kr)]e^{i(kr-wt)} \quad (1)$$

$$E_\theta = \frac{\rho \, Sin(\theta)}{4\pi\varepsilon_o \, r^3}\left[\{1 - (kr)^2\} - i(kr)\right]e^{i(kr-wt)} \quad (2)$$

$$H_\phi = \frac{\omega\rho Sin(\theta)}{4\pi \, r^2}[-kr - i]e^{i(kr-wt)} \quad (3)$$

Figure 1: Spherical co-ordinate system used to analyze electric dipole and resulting EM field solutions

Alternatively the electric dipole solution can be expressed as a superposition of sinusoidal waves which propagate at the speed of light. Using the identity: $e^{i(kr-\omega t)} = Cos(kr-\omega t) + i\, Sin(kr-\omega t)$ and extracting the imaginary part of the solution yields:

$$E_r = \frac{\rho\, Cos(\theta)}{2\pi\varepsilon_o\, r^3}\left[Sin(kr-\omega t) - (kr)\, Cos(kr-\omega t)\right] \qquad (4)$$

$$E_\theta = \frac{\rho\, Sin(\theta)}{4\pi\varepsilon_o\, r^3}\left[Sin(kr-\omega t) - (kr)^2\, Sin(kr-\omega t) - (kr)\, Cos(kr-\omega t)\right] \qquad (5)$$

$$H_\phi = \frac{\omega\rho\, Sin(\theta)}{4\pi\, r^2}\left[-(kr)\, Sin(kr-\omega t) - Cos(kr-\omega t)\right] \qquad (6)$$

It should be noted that all of the above solutions are only valid for distances (r) much greater than the dipole length ($d_o$). In the region next to the source (r ~ $d_o$), the source cannot be modelled as a sinusoid: $Sin(\omega t)$. Instead it must be modelled as a sinusoid inside a Dirac delta function: $\delta[r - d_o Sin(\omega t)]$. The solution of this hyper-near-field problem can be calculated using the Liénard-Wiechert potentials [3, 4, 5]

## 2.2 Analysis of instantaneous phase speed and group speed

It is noted from the above analysis that the field solutions of the electric dipole can be written as a sum of sinusoidal waves, which travel away from the dipole source at the speed of light. Even if the waves are generated by unique physical mechanisms, only the superposition of the waves is observable at any point in space. These wave components in effect form a new wave which may have different properties than the original components. Only the longitudinal and transverse wave components are real since they can be decoupled by proper configuration of a measurement antenna. The following analysis derives general relations that are used to determine the instantaneous phase and group speed vs distance graphs for the longitudinal and transverse field components.

### 2.2.1 Derivation of phase speed relation

In this section a mathematical relation is derived which enables the instantaneous phase speed of a wave to be determined from its phase vs distance curve. Given a propagating wave of the form: $Sin(kr-\omega t)$ the instantaneous wave phase speed ($c_{ph} = \Delta r/\Delta t$) is the propagation speed of a point of constant phase ($\Delta\theta = \omega\,\Delta t$) on the wave. Solving this relation for time ($\Delta t = \Delta\theta/\omega$) and inserting it into the phase speed relation yields: $c_{ph} = \omega/(\Delta\theta/\Delta r)$. In the limit and using the relation ($\omega = c_o k$, where k is a far-field constant) the instantaneous phase speed becomes [6]:

$$c_{ph} = \omega \bigg/ \frac{\partial\theta}{\partial r} = c_o k \bigg/ \frac{\partial\theta}{\partial r} \qquad (7)$$

Alternatively this phase speed relation can be derived from the known relation: $c_{ph} = \omega/k$. Solving the phase ($\Delta\theta = k\,\Delta r$) for (k) and inserting it in the phase speed equation yields: $c_{ph} = \omega/(\Delta\theta/\Delta r)$. In the limit this becomes Eqn. 7. Since $\omega = c_o k$ (where k is a far-field constant) the phase speed becomes: $c_{ph} = (c_o k)/(\Delta\theta/\Delta r) = c_o / [\Delta\theta/\Delta(kr)]$.

Inserting the relation: $k = 2\pi/\lambda$ yields: $c_{ph} = 2\pi c_o / [\Delta\theta/(\Delta(r_{el}))]$. In the limit the instantaneous phase speed relation becomes:

$$c_{ph} = \left.\frac{360\, c_o}{\dfrac{\partial \theta}{\partial r_{el}}}\right|_{\theta \text{ in deg}} \qquad (8)$$

where the electrical length is: $r_{el} = r/\lambda$. A more rigorous derivation of this relation can be found in a previous paper by the author [3, 4, 7]. The above relation (ref. Eq. 8) indicates that the instantaneous phase speed is inversely proportional to the slope of the phase vs distance curve. Note that zero slope on this curve would indicate an infinite instantaneous phase speed.

### 2.2.2 Derivation of group speed relation

In this section a mathematical relation is derived that enables the instantaneous group speed of a wave to be determined given its phase vs distance curve. The group speed is known to be the speed at which wave energy and information travel. It can be calculated by considering two Fourier components of a wave group which form an amplitude modulated signal: $Sin(\omega_1 t - k_1 r) + Sin(\omega_2 t - k_2 r) = Sin(\Delta\omega t - \Delta k r)\, Sin(\omega t - kr)$ in which: $\Delta\omega = (\omega_1 - \omega_2)/2$, $\Delta k = (k_1 - k_2)/2$, $\omega = (\omega_1 + \omega_2)/2$, and $k = (k_1 + k_2)/2$. The instantaneous group speed ($c_g = \Delta\omega/\Delta k$) is the propagation speed of a point of constant phase ($\Delta\theta = \Delta k\, \Delta r$) of the modulation component of the modulated wave. Solving the phase relation for ($\Delta k$) and inserting it into the group speed relation yields: $c_g = \Delta\omega/(\Delta\theta/\Delta r) = [(\Delta\theta/(\Delta r\, \Delta\omega)]^{-1}$. In the limit and using the relation ($\omega = c_o k$) the instantaneous group speed becomes [6]:

$$c_g = \left[\frac{\partial^2 \theta}{\partial r \partial \omega}\right]^{-1} = \frac{1}{c_o}\left[\frac{\partial^2 \theta}{\partial r \partial k}\right]^{-1} \qquad (9)$$

A more rigorous derivation of this relation can be found in a previous paper by the author [3, 4, 7]. The above relation can also be made a function of (kr) by multiplying the numerator and the denominator by (k) and using the relation ($k = w/c_o$) yielding: $c_g = c_o\, [\omega \Delta\theta / (\Delta\omega \Delta kr)]^{-1}$. In the limit this becomes: $c_g = c_o\, [(d/d\omega)\, \omega\, d\theta/d(kr)]^{-1} = c_o\, [\omega\, (d/d\omega)\, (d\theta/d(kr)) + d\theta/d(kr)]^{-1}$. Using the relation ($\omega = c_o k$) the instantaneous group speed becomes: $c_g = c_o\, [kr\, (d/d(kr))\, (d\theta/d(kr)) + d\theta/d(kr)]^{-1}$. Using the relation for the electrical wavelength ($r_{el} = r/\lambda = kr/(2\pi)$), the group speed becomes: $c_g = 2\pi\, c_o\, [r_{el}\, (d/d(r_{el}))\, (d\theta/d(r_{el})) + d\theta/\, d(r_{el})]^{-1}\,|_{\theta \text{ in rad.}}$ In conclusion the instantaneous group speed becomes:

$$c_g = \left.\frac{360\, c_o}{\left[r_{el}\, \dfrac{\partial^2 \theta}{\partial r_{el}^2} + \dfrac{\partial \theta}{\partial r_{el}}\right]}\right|_{\theta \text{ in deg}} \qquad (10)$$

The above relation (ref. Eq. 10) indicates that the instantaneous group speed is inversely proportional to both the curvature and the slope of the phase vs distance curve. Note that if the denominator of the above equation (ref. Eq. 10) is zero, the group speed will be infinite. Also note that if the curvature is zero, the group speed equation (ref. Eq. 10) will be the same as the phase speed equation (ref. Eq. 8).

### 2.2.3 Radial electric field (E_r)

Applying the above phase and group speed relations (ref. Eq: 7, 9) to the radial electrical field (E_r) component (ref. Eq. 1 or 4) yields the following results [7]:

$$\theta = kr - Tan^{-1}(kr) \underset{kr \ll 1}{\approx} -\frac{1}{3}(kr)^3 \quad (11)$$

$$c_{ph} = c_o\left(1 + \frac{1}{(kr)^2}\right) \underset{kr \ll 1}{\approx} \frac{c_o}{(kr)^2} \underset{kr \gg 1}{\approx} c_o \quad (12)$$

$$c_g = \frac{c_o\left(1 + (kr)^2\right)^2}{3(kr)^2 + (kr)^4} \underset{kr \ll 1}{\approx} \frac{c_{ph}}{3} \underset{kr \gg 1}{\approx} c_o \quad (13)$$

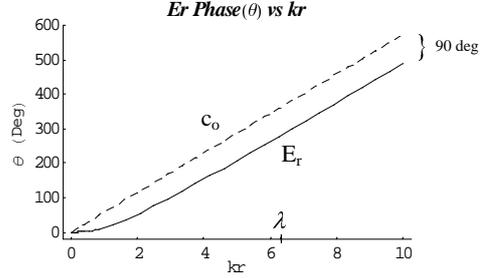

Figure 2: E_r Phase vs kr

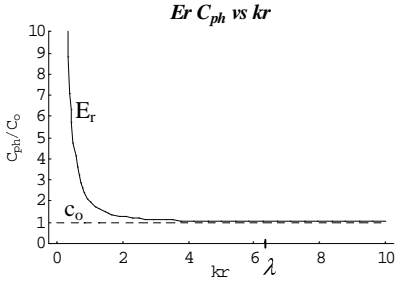

Figure 3: E_r c_ph vs kr

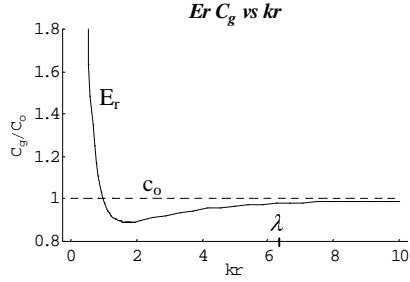

Figure 4: E_r c_g vs kr

### 2.2.4 Transverse electric field (E_θ)

Applying the above phase and group speed relations (ref. Eq. 7, 9) to the transverse electrical field (E_θ) component (ref. Eq. 2 or 5) yields the following results [7]:

$$\theta = kr - Cos^{-1}\left(\frac{1-(kr)^2}{\sqrt{1-(kr)^2+(kr)^4}}\right) \quad (14)$$

$$c_{ph} = c_o\left(\frac{1-(kr)^2+(kr)^4}{-2(kr)^2+(kr)^4}\right) \quad (15)$$

$$c_g = \frac{c_o\left(1-(kr)^2+(kr)^4\right)^2}{-6(kr)^2+7(kr)^4-(kr)^6+(kr)^8} \quad (16)$$

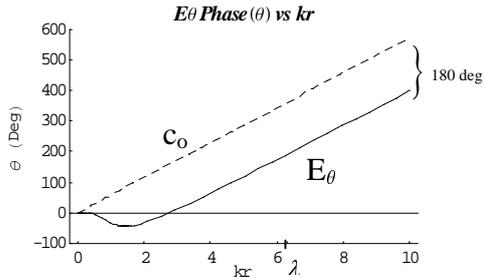

Figure 5: E_θ phase (θ) vs kr

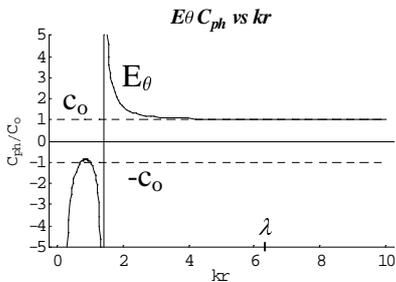

Figure 6: E_θ c_ph vs kr

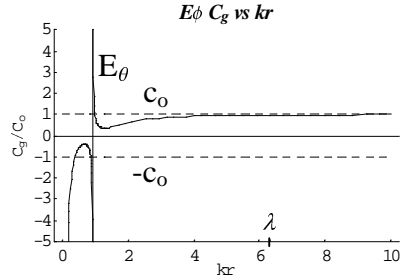

Figure 7: E_θ c_g vs kr

### 2.2.5 Transverse magnetic field ($H_\phi$)

Applying the above phase and group speed relations (ref. Eq. 7, 9) to the transverse magnetic field ($H_\phi$) component (ref. Eq. 3 or 6) yields the following results [7]:

$$\theta = kr - Cos^{-1}\left(\frac{-kr}{\sqrt{1+(kr)^2}}\right) \quad (17)$$

$$c_{ph} = c_o\left(1 + \frac{1}{(kr)^2}\right) \quad (18)$$

$$c_g = \frac{c_o\left(1+(kr)^2\right)^2}{3(kr)^2 + (kr)^4} \quad (19)$$

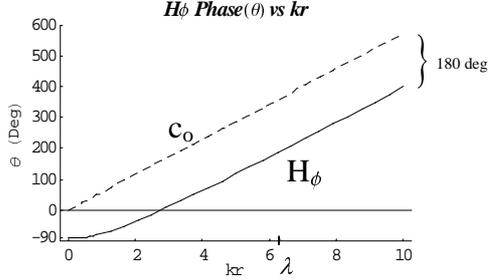

Figure 8: $H_\phi$ phase ($\theta$) vs kr

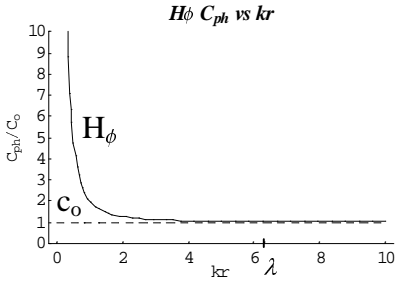

Figure 9: $H_\phi$ $c_{ph}$ vs kr

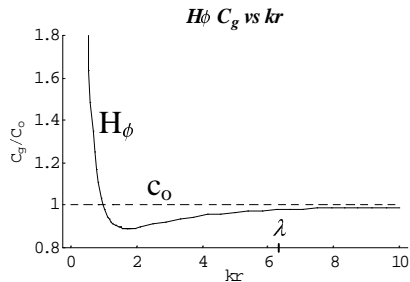

Figure 10: $H_\phi$ $c_g$ vs kr

### 2.2.6 Animated field plots

In this section, animated contour plots are presented which show how the longitudinal and transverse electric fields propagate. A cosinusoidal dipole source is used and the resultant fields are assumed to be a vectoral sum of all the wave components. The resultant field magnitude and phase are then inserted into a propagating cosine wave: Mag Cos($\omega$t + ph) and plotted at different moments in time. The plots are generated using Mathematica Ver. 3 software. The code generates 24 plots evenly spaced within a specified analysis period. Several of the resultant frames are shown below. The vertical dipole source is located in the center of the plots. The frames shown below (ref. Fig. 11) are animated plots of the longitudinal electric field. They clearly show that the waves are generated at the source and propagate away from the source.

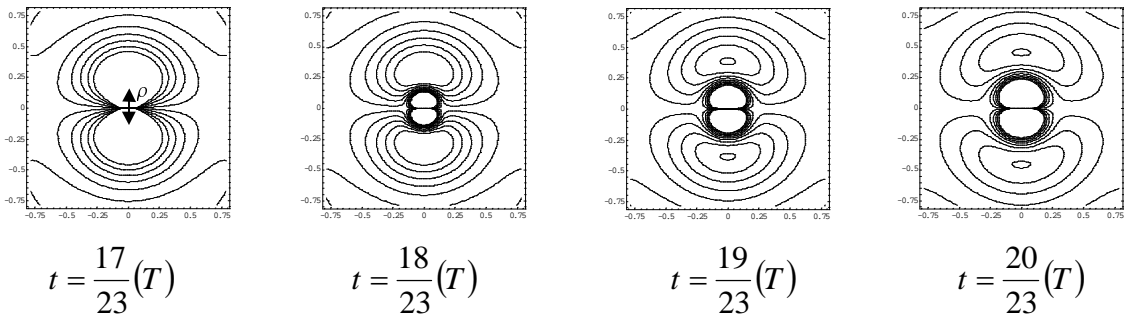

$t = \frac{17}{23}(T)$  $\quad t = \frac{18}{23}(T)$  $\quad t = \frac{19}{23}(T)$  $\quad t = \frac{20}{23}(T)$

## Mathematica code used to generate animations

```
Eth=MagEth*Cos[w*t+PhEth];
MagEth=po/4/Pi/eo*Sqrt[(1-(k*r)^2)^2+(k*r)^2]/r^3*Sin[th];
PhEth=-k*r+ArcCos[(1-(k*r)^2)/Sqrt[1-(k*r)^2+(k*r)^4]];
Er=MagEr*Cos[w*t+PhEr];
MagEr=po/2/Pi/eo*Sqrt[1+(k*r)^2]/r^3*Cos[th];
PhEr=-k*r+ArcTan[k*r];
L=1;k:=2*3.14159/L;c=3*10^8;w=2*3.141159*c/L;
T=L/c;po=1.6*10^(-19);eo=8.85*10^(-12);
r=Sqrt[x^2+y^2];
Animate[ContourPlot[Er/(1*10^(-7)),{x,-Pi/4,Pi/4},{y,-Pi/4,Pi/4},
PlotPoints->100],{t,0,1*T},ContourShading->False,
Contours->{-.9,-.7,-.5,-.3,-.1,.1,.3,.5,.7,.9}]
```

Figure 11: $E_r$ near-field wave animation plots

The frames shown below (ref. Fig. 12) are animated plots of the transverse electrical field. The plots clearly show that the waves are created outside the source and propagate toward and away from the source.

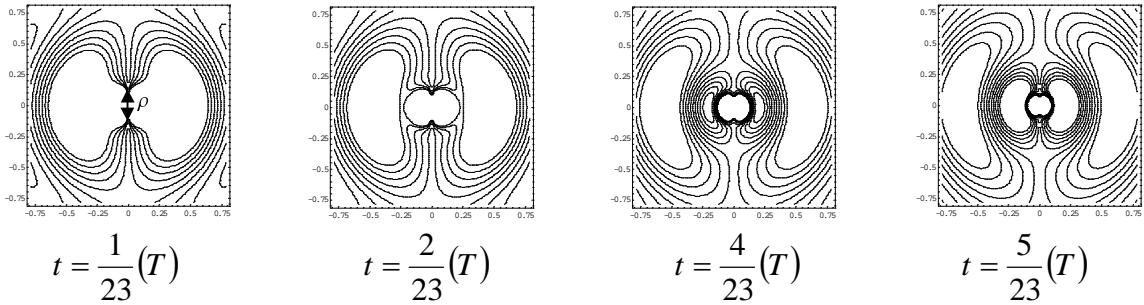

$t = \dfrac{1}{23}(T)$  $\quad\quad$  $t = \dfrac{2}{23}(T)$  $\quad\quad$  $t = \dfrac{4}{23}(T)$  $\quad\quad$  $t = \dfrac{5}{23}(T)$

Figure 12: $E_\theta$ near-field wave animation plots

The frames shown below (ref. Fig. 13) are animated plots of the longitudinal and transverse electrical fields vectorially added together (vector plot). The vertical dipole source is located in the middle of the left-hand side of the plot

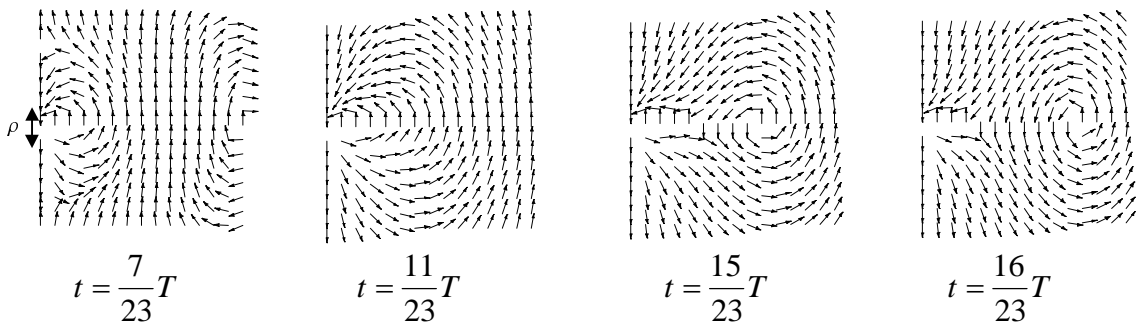

$t = \dfrac{7}{23}T$  $\quad\quad$  $t = \dfrac{11}{23}T$  $\quad\quad$  $t = \dfrac{15}{23}T$  $\quad\quad$  $t = \dfrac{16}{23}T$

## Additional Mathematica code used to generate vector plot animations
(Add this code to the previous code used above)

```
Ex=Er*Sin[th]+Eth*Cos[th];
Ey=Er*Cos[th]-Eth*Sin[th];
r=Sqrt[x^2+y^2];
th=ArcCos[y/(Sqrt[x^2+y^2])];
<<Calculus`VectorAnalysis`
<<Graphics`PlotField`
Ett={Ex,Ey};
Etmag=Sqrt[Ex^2+Ey^2];
Animate[PlotVectorField[Ett/Etmag,{x,0,0.5},{y,-0.25,0.25}],{t,0,T}]
```

Figure 13: Animated plot of Er and $E_\theta$ vectorially added together

A more detailed plot of the total electric field can be obtained by using the fact that a line element crossed with an electric field is zero. The resulting partial differential equation can be solved yielding [1, 2]:

Resultant Equation: $\quad \dfrac{\partial}{\partial r}\left(rC_\phi Sin\theta\right)dr + \dfrac{\partial}{\partial \theta}\left(rC_\phi Sin\theta\right)d\theta = 0 \quad$ (20)

Solution: $\quad \sqrt{1 + \dfrac{1}{(kr)^2}}\, Sin^2\theta\, Cos\left[kr - Tan^{-1}(kr) - \omega t\right] = Const \quad$ (21)

A contour plot of this solution yields the plot below (ref. Fig.14).

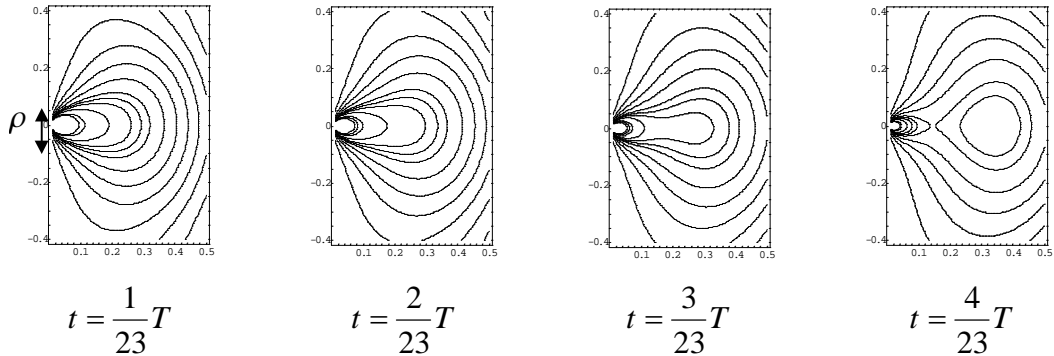

$t = \dfrac{1}{23}T \qquad t = \dfrac{2}{23}T \qquad t = \dfrac{3}{23}T \qquad t = \dfrac{4}{23}T$

Mathematica code used to generate E field contour plot

```
L= 1;c=3*10^8; f= c/L;T=1/f;w=2*N[Pi]*f;k= w/c;
fn= Sqrt[1/ (k*r)^2+ 1] *Cos [Th]^2* Cos[w*t-k*r+ArcTan[k*r]] ;
r = Sqrt[x^2 + y^2] ;Th = ArcTan[y/x] ;
Animate[ContourPlot[fn, {x, 0.01,.5}, (y, -.4,.4), PlotPoints-> 100] ,
{t, 0, T} , ContourShading->False,
Contours->{-. 9, -.7, -.5, -.3, -.1, .1, .3, .5, .7, .9},
AspectRatio->3/2]
```

Figure 14: Animated plot of E field in near-field

Further away from the source the plot of the electric field becomes:

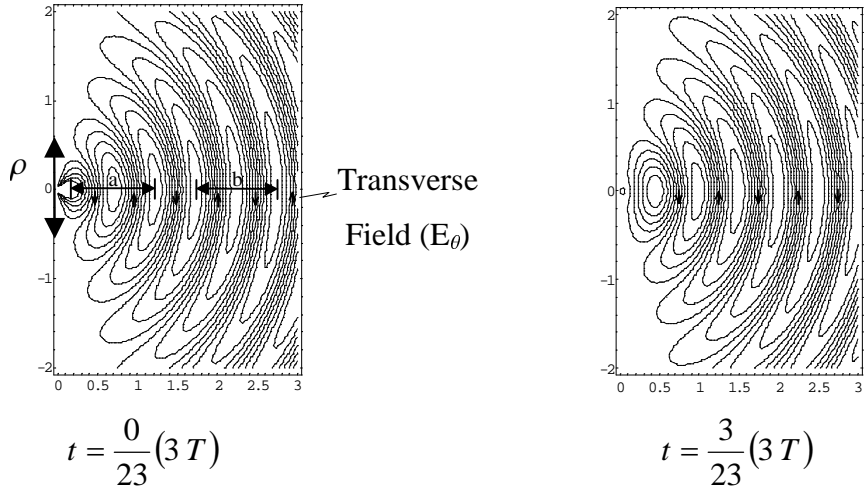

Figure 15: Plot of E field in far-field

Note that careful inspection of the plot reveals that the wavelength of the transverse electric field in the near-field (a) is larger than the wavelength in the far-field (b). The phase speed ($c_{ph}$) is known to be a function of wavelength ($\lambda$) and frequency (f): $c_{ph} = \lambda\, f$. Solving the relation for (f), which is constant both in the near-field and far-field, yields: $f = Cph_{near}/\lambda_{near} = Cph_{far}/\lambda_{far}$. Solving this for $Cph_{near}$ yields: $Cph_{near} = Cph_{far}\,(\lambda_{near}/\lambda_{far})$. Since $\lambda_{near} > \lambda_{far}$ the phase speed of the transverse electric field is larger than the speed of light. ($c_{ph} > c_o$).

### 2.2.7 Interpretation of theoretical results

The above theoretical results suggest that longitudinal electric field waves and transverse magnetic field waves are generated at the dipole source and propagate away. Upon creation, the waves (phase and group) travel with infinite speed and then rapidly reduce to the speed of light after they propagate about one wavelength away from the source. In addition, transverse electric field waves (phase and group) are generated approximately one-quarter wavelength outside the source and propagate toward and away from the source. Upon creation, the transverse waves travel with infinite speed. The outgoing transverse waves reduce to the speed of light after they propagate about one wavelength away from the source. The inward propagating transverse fields rapidly reduce to the speed of light and then rapidly increase to infinite speed as they travel into the source. In addition, the above results show that the transverse electrical field waves are generated about 90 degrees out of phase with respect to the longitudinal waves. In the near-field the outward propagating longitudinal waves and the inward propagating transverse waves combine together to form a type of oscillating standing wave. Note that unlike a typical standing wave the the outward and inward waves are completley different types of waves (longitudinal vs transverse) and can be separated by proper orientation of a detecting antenna. In addition, it should also be noted that both the phase and group waves are not confined to one side of the speed of light boundary and propagate at speeds above and below the speed of light in specific regions from the source.

The mechanism by which the electromagntetic near-field waves become superluminal can be understood by noting that the field componets can be considered rectangular vector components of the total field (ref. Fig. 16). For example, the vector diagram for the longitudinal electric field is (ref. Eq. 4):

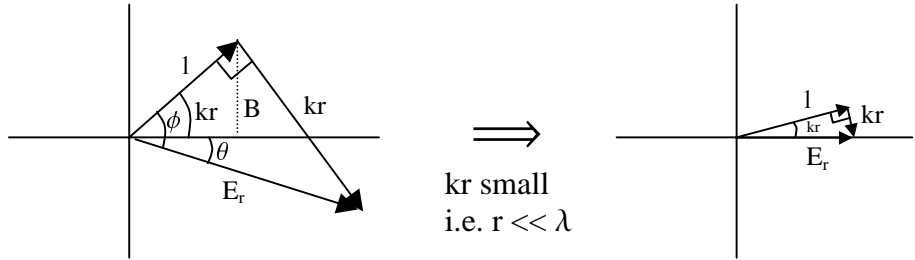

Figure 16: Vector diagram for longitudinal electric field

From this vector diagram it can be seen that the phase of the longitudinal electric field is: $\theta = \phi - kr$. Also it can be seen that angle: $\phi = \text{ArcTan}[kr]$. Combining these relations yields phase relation Eq. 11: $\theta = \text{ArcTan}[kr] - kr$. Note that for small ($kr \therefore r \ll \lambda$) the angle bisector: $B = 1 \sin(kr) \cong kr$ has about the same length as vector ($kr$). Therefore when the ($kr$) is small the two vector components add together to form a longitudinal electric field vector which has nearly zero phase. Note that the angle bisector approximation is valid for several values of ($kr$) when ($kr$) is small. This result can also be seen by Taylor expanding the phase relation for small ($kr$) yielding: $\theta = kr - [kr + (kr)^3/3 + O(kr)^5] = (kr)^3/3 + O(kr)^5$, where $kr = \omega r/c$. These results show that very near the dipole source the phase of the longitudinal electric field is zero, causing both the phase speed and the group speed to be infinite (ref Eq. 7, 9). In the near-field the phase increases to $(kr)^3/3$, causing the phase speed to be: $c_o/(kr)^2$ (ref. Eq. 7) and the group speed to be: $c_o/(kr)^2/3$ (ref. Eq. 9). In the far-field the phase becomes: $\pi - kr$, causing both the phase speed and the group speed to be equal to the speed of light (ref Eq. 7, 9). The other components of the electromagnetic field ($E_\theta$, $H_\phi$) can also be analysed in the same way yielding similar results.

## 3 Experimental results

### *3.1 Description of experiment*

A simple experimental setup (ref. Fig. 17) has been developed to qualitatively verify the transverse electric field phase vs distance plot predicted from standard EM theory (ref. Fig. 5).

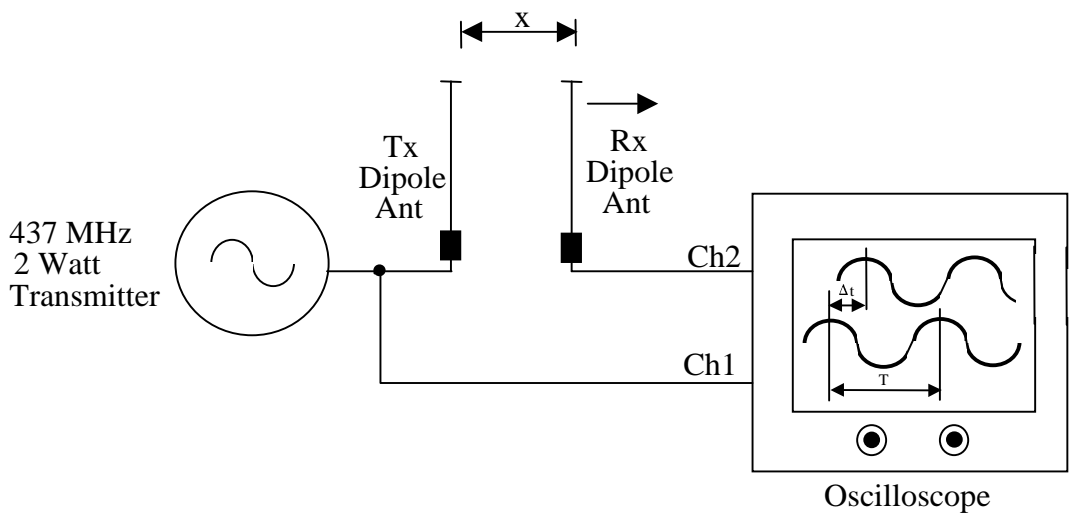

Figure 17: Experimental setup

The experiment setup consists of a high frequency UHF FM transmitter (Hamtronics model no. TA451)[1] which generates a 437MHz (68.65cm wavelength), 2 watt sinusoidal electrical signal. The output of the transmitter is connected with a RG58 coaxial cable to a vertical dipole antenna designed for the carrier frequency (model no. RA3126)[2]. The output of the transmitter is also connected to channel 1 of the input of a high frequency 500MHz digital oscilloscope (model no. HP54615B). The transmitter output, cable, antenna, and oscilloscope input all have 50 Ohm impedance in order to minimize reflections. A second identical receiver dipole antenna is connected to channel 2 of the high frequency oscilloscope and the antenna is positioned parallel to the vertical transmitting antenna. The sinusoidal signals from the two antennas are monitored with the oscilloscope, triggered to channel 1. The phase difference between the signals is measured using the oscilloscope measurement cursors as the antennas are moved apart from 5 cm to 70 cm in increments of 5 cm (measurements made with a ruler). The oscilloscope calculates the phase from the measured time delay ($\Delta t$) and the measured wave period (T): $\theta_{deg} = (360\Delta t)/T$. The phase vs distance data is analyzed using HPVEE (Ver. 4.01) PC software. The data is then curvefit with a 3$^{rd}$ order polynomial and the data is superimposed to visually verify the accuracy of the curvefit. The phase speed vs distance curve and the group speed vs distance curve are then generated by differentiating the resultant curvefit equation with respect to space and using the transformation relations (ref. Eq. 8, 10).

---

[1] Ref. Internet site: www.hamtronics.com
[2] Ref. Internet site: www.elfa.se - part no. 78-069-95

## 3.2 Experimental results

The following graph (ref. Fig. 19) is a plot of the phase vs distance data [ref. Fig. 18] taken during one experiment. The phase and group speed graphs were generated by curvefitting the experimental data and inserting the curvefit equation into the phase and group speed transformations: (ref. Eq. 8, 10). The first data point is not real and was added to improve the polynomial curvefit. The curvefit yielded the following polynomial: $ph = (132.2) + (-262.5)r_{el} + (838.9)r_{el}^2 + (-353.4)r_{el}^3$

| Data # | $r_{el}$ (cm) | Ph (Deg) |
|---|---|---|
| 0 | 0 | 140.0 |
| 1 | 5 | 111.7 |
| 2 | 10 | 102.4 |
| 3 | 15 | 108.6 |
| 4 | 20 | 121.0 |
| 5 | 25 | 136.6 |
| 6 | 30 | 155.2 |
| 7 | 35 | 170.7 |
| 8 | 40 | 195.5 |
| 9 | 45 | 211.0 |
| 10 | 50 | 245.2 |
| 11 | 55 | 282.4 |
| 12 | 60 | 316.6 |
| 13 | 65 | 325.9 |
| 14 | 70 | 366.2 |

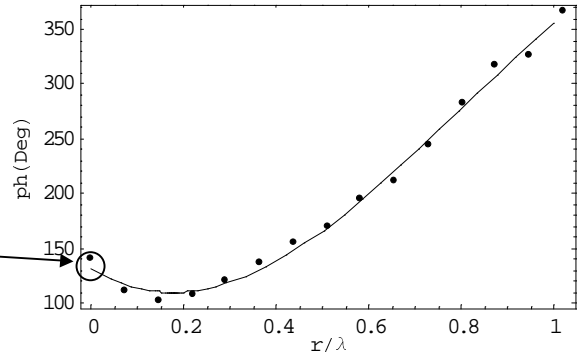

Figure 18: Phase data vs $r_{el}$

Figure 19: Curvefit of phase vs distance data ($r_{el}$)

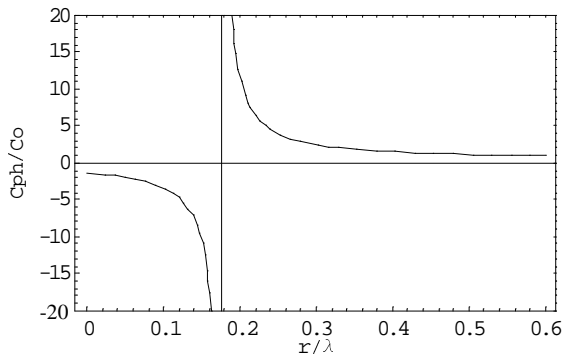

Figure 20
Calculated phase speed ($c_{ph}/c_o$) vs distance ($r_{el}$) graph

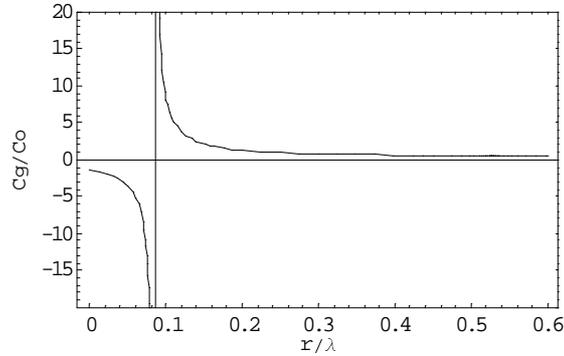

Figure 21
Calculated group speed ($c_g/c_o$) vs distance ($r_{el}$) graph

It should be noted that these experimental results are only qualitative due to EM reflections from nearby walls and objects. Quantitative measurements can only be attained in an anechoic chamber. The experiment has been repeated several times in different parts of a 4 x 4m (area) x 2m (height) room at different angular orientations to the walls and the phase vs distance curve always appears the same within 10%. It is also observed that changing the scope input impedance from 50 Ohms to 1M Ohm input impedance does not noticeably affect the phase vs distance curve. Since no effect is observed it is concluded that the Tx antenna to Rx antenna variable capacitance combined with the scope input impedance (thereby forming a high pass filter) is not the cause of the phase change. Experimentally it is observed that the electrical field near the source (less than 0.6 $\lambda$) is at least an order of magnitude greater than electric field several wavelengths away from the source, which may be reflected. It is concluded that the observed field near the source is predominantly due to near-field effects thereby making the observed results qualitatively reliable. The experimental results (ref. Fig. 19, 20, 21) are qualitatively similar to the electric dipole solution presented (ref. Fig. 5, 6, 7). Differences between experiment and the theory presented can be attributed to EM reflections and also to the fact that the theoretical model for a real dipole antenna is somewhat different from the simple electric dipole solution presented.

### *3.3 Interpretation of experimental results*

Analysis of the experimentally derived phase vs distance curve (ref. Fig. 19) indicates that the phase vs distance curve generated from the experimental data is very similar to the curve predicted from electric dipole theory (ref. Fig. 5). Performing the experiment in an anacroic chamber and improving theoretical model for the dipole antenna should yield a better match between theory and experiment. The phase speed (ref. Fig. 20, 6) and group speed (ref. Fig. 21, 7) vs distance curves do not match theory as well as the phase vs distance curves (ref. Fig. 19, 5). This difference is due to the fact that small errors in the experimental data and curvefit become magnified after differentiating the data, which is required by the phase and group speed transformation relations (ref. Eq. 8, 10). Although the experimental results are not as accurate as they can be, it can be qualitatively seen from the results that transverse electric field waves (phase and group) are generated approximately one-quarter wavelength outside the source and propagate toward and away from the source (ref. Fig. 20, 21). Upon creation, the transverse waves travel with infinite speed. The outgoing transverse waves reduce to the speed of light after they propagate about one wavelength away from the source. Note that infinite phase speed and group speed is expected since the phase vs distance curve (ref. Fig. 19) has a minimum at about one quarter wavelength distance from the source (ref. Eq. 8, 10).

## 4 Discussion

### *4.1 Arguments against superluminal interpretation*

#### 4.1.1 Superluminal illusion due to velocity dependent field cancellation

An argument against the superluminal interpretation presented in this paper has appeared in the literature [8, 9, 10] in which it is argued that the electrical field can be decomposed into 3 terms: a retarded coulomb field component, a retarded velocity dependent component, and a retarded acceleration component.

$$E = \frac{-q}{4\pi\varepsilon_o}\left[\frac{e_{r'}}{r'^2} + \frac{r'}{c}\frac{d}{dt}\left(\frac{e_{r'}}{r'^2}\right) + \frac{1}{c^2}\frac{d^2}{dt^2}e_{r'}\right] \qquad (22)$$

Where (r´) denotes the retarded distance, (q) charge, (t) time, (c) speed of light, and ($e_{r'}$) unit vector from retarded position to field point. Note, retardation refers to the time it takes for the field to propagate a distance (r) at the speed of light. In the near-field the first two terms dominate and the velocity component partially cancels the effects of the retardation causing the field to be nearly instantaneous in the near-field. In the far-field the retarded acceleration component dominates. The above equation can be used to analyze the electrical field produced by an oscillating dipole. According to Feynman [8] the electric field for the electric dipole becomes:

$$E = \frac{-1}{4\pi\varepsilon_o r^3}\left[-p^* - \frac{3(p^* \cdot r)r}{r^2} + \frac{1}{c^2}\{\ddot{p}(t-r/c) \times r\} \times r\right] \quad (23)$$

$$\text{where } p^* = p(t-r/c) + \frac{r}{c}\dot{p}(t-r/c)$$

Given the retarded dipole moment {p=$\rho$ Sin(kr-$\omega$t)}, and using the relation (kr = $\omega$r/c) the longitudinal component of the electric field becomes:

$$E_r = \frac{Cos(\theta)}{2\pi\varepsilon_o r^3}\left[p + \frac{r}{c}\dot{p}\right] = \frac{\rho\, Cos(\theta)}{2\pi\varepsilon_o r^3}\left[Sin(kr-\omega t) - (kr)\,Cos(kr-\omega t)\right] \quad (24)$$

The above solution is equivalent to the longitudinal electrical field solution presented at the beginning of the paper (ref. Eq. 1, 4), which yields the superluminal phase and group speed curves (ref. Fig. 3, 4). This result shows that the solution is the sum of two retarded longitudinal vectorial components resulting in a superluminal longitudinal electrical field. The first sinusoidal term is due to the retarded coulomb field and the second cosinusoidal term is due to the retarded velocity dependent component. Opponents would argue that the two retarded longitudinal components are real and that they combine together to produce a superluminal illusion. The interpretation presented in this paper is that the resultant longitudinal electrical field is a superposition of all the retarded wave components. At each point in space the wave components vectorially add together to form a new wave that propagates superluminally as it is created and reduces to the speed of light after it has propagated approximately one wavelength from the source.

In addition, the transverse component of the electric field can also be determined using (ref. Eq. 23) yielding:

$$E_\theta = \frac{Sin(\theta)}{4\pi\varepsilon_o r^3}\left[-p - \frac{r}{c}\dot{p} + \frac{1}{c^2}\ddot{p}\right]$$

$$= \frac{\rho\, Sin(\theta)}{4\pi\varepsilon_o r^3}\left[Sin(kr-\omega t) - (kr)^2 Sin(kr-\omega t) - (kr)\,Cos(kr-\omega t)\right] \quad (25)$$

The solution is also equivalent to the longitudinal electrical field solution presented at the beginning of the paper (ref. Eq. 2, 5), which yields the superluminal phase and group speed curves (ref. Fig. 6, 7). Opponents would also argue that the transverse electrical field is composed of three retarded sinusoidal propagating transverse waves which combine together to form a superluminal illusion. It is argued in this paper that these waves add vectorially as they propagate from the source and form a new type of wave that has different properties from its components. An interesting proof of this is that it is known (by using the pointing vector) that only the far-field transverse components ($E_\theta$, $H_\phi$) radiate, and that the near-field components form a type of standing wave and do not radiate [2]. But it can be seen from the above solutions (ref. Eq. 24, 25) that all of the components of the electric field propagate away from the dipole source and none of the components propagate toward the source which is required for the near-field propagating fields to not radiate from the source. Using the interpretation presented in this paper it can be understood that in the near-field, the outward propagating longitudinal waves and the inward propagating transverse waves combine together in the near-field and form a type of oscillating standing wave. In the far-field only the outward propagating transverse waves radiate.

### 4.1.2 Superluminal illusion due to presence of standing waves

It is also suggested by some authors that the near-field of an electrical dipole consists of an electrical field which grows and collapses synchronized with the oscillation of the electric dipole, resulting in a type of standing wave. Since standing waves are thought to be the addition of transmitted and reflected waves the resultant field may yield phase shifts unrelated to how the fields propagate, thereby refuting the results presented in this paper. It is argued that the theory and experimental results presented in this paper suggest that the oscillating electrical dipole generates an electrical field  composed of two types of waves: longitudinal and transverse. The longitudinal electrical wave is generated at the source and propagates superluminally away from the source. The transverse electrical wave is generated about one quarter wavelength outside the source and propagates superluminally toward and away from the source. The apparent growth and collapse of the near-field electrical field is due to the fact that the waves are produced 90 degrees out of phase. In the near-field the outward propagating longitudinal waves and the inward propagating transverse waves combine together to form a type of oscillating standing wave. Note that unlike a typical standing wave the outward and inward waves are completely different types of waves (longitudinal vs transverse) and can be separated by proper orientation of a detecting antenna.

### *4.2 Causality issue*

Although superluminal phase speeds are known to exist in other physical systems (eg. EM wave propagation in the ionosphere [11] ), group speeds exceeding the speed of light are not known to exist. In Einstein's 1907 paper [12] he indicated that although relativity does not prohibit the existence of superluminal signals (group speed > c) relativity does predict that superluminal signals can be seen by a moving observer to travel backward in time. Einstein concluded that a superluminal signal (w) propagating a known distance (l) would be seen by a moving observer (v) to have crossed the distance in time: $\Delta t = [(1-w\,v/c^2)/(w-v)]\,l$. If the signal speed is hyperluminal: $w > (c^2)/v$ then the signal would be seen by the moving observer to travel backward in time ($\Delta t$ becomes negative). The result can also be derived from the relativistic equation for time: $\Delta t' = \gamma\,[\Delta t - (v/c^2)\,\Delta x]$, since $\Delta t = l/w$ and $\Delta x = l$ then: $\Delta t' = \gamma\,(l/w - v\,l/c^2) < 0$ for time reversal. Solving for (w) yields: $w > (c^2)/v$. This effect can also be intuitively understood by using a spacetime diagram, with the moving coordinates $(xc', t')$ superimposed on the reference frame of a stationary observer $(xc, t)$. [13]

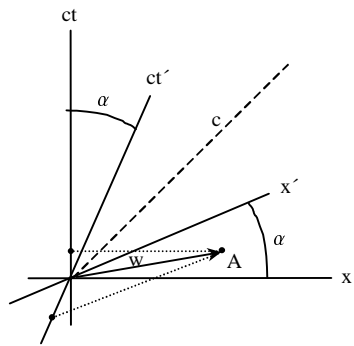

Figure 22. Spacetime diagram showing a mechanism for time reversal

The (x´) and (t´) axes are at angles ($\alpha$) with respect to the (x) and (t) axes, where $\alpha$ = ArcTan(v/c). If a signal is transmitted superluminally (with respect to the stationary reference frame) from the origin to point (A), then the signal speed is: ct/x < Tan($\alpha$), but ct/x = c/w, therefore c/w < v/c. Solving this relation for (w) yields: $w > c^2/v$. Although ($\Delta$t) with respect to the stationary reference frame is positive, ($\Delta$t) with respect to the moving reference frame is negative, indicating that from the moving reference frame the signal will be seen to travel backward in time. This is commonly referred to as a violation of causality where effect precedes cause. Although special relativity does not forbid that signals can travel faster than the speed of light, it does predict that if signals travel hyperluminally ($w > c^2/v$), the signal would be seen by a moving observer to travel backward in time. From the theory presented in this paper, it is seen that all of the waves generated by an oscillating electric dipole travel with infinite speed at their point of creation and travel superluminally within a limited region of space (~0.1 $\lambda$). It should be noted that this region of space can be very large for low frequencies (frequencies less than 30MHz yield: 0.1 $\lambda$ > 1m). Therefore, it is concluded that according to relativity theory a moving observer can see these superluminally propagating waves propagating backward in time provided $w > c^2/v$. It should be noted that the moving reference frame can travel subluminally.

### *4.3 Speed of information propagation and detection*

Although the speed of information propagation (group speed ($c_g$)) may be superluminal, the speed of information propagation and detection may be less. If a sinusoidally modulated signal propagates with a group speed ($c_g$) and the sinusoidal modulation (Period T = 1/f) propagates a distance (d) in time (t), detection of the signal may require several cycles (nT) of the signal in order to decode the information. The speed of information detection ($c_{inf}$) can then be modelled: $c_{inf}$ = d / (t+nT). Since d = $c_g$ t, then $c_{inf}$ = ($c_g$ t) / (t+nT). In the far-field the propagation time (t) can be much larger than the number of cycles (nT) needed to decode the signal, therefore: $c_{inf}$ = $c_g$. In the near-field the propagation time (t) can be much smaller than the number of cycles (nT) needed to decode the signal, therefore: $c_{inf}$ = $c_g$ t / (nT). This result shows that depending on the number of cycles required to detect the signal, the speed of information propagation and detection may be significantly less than the group speed in the near-field. It is known from Fourier theory that several cycles of a sinusoid are required for the information (frequency) to be determined. Therefore, if information detection is based on Fourier decomposition of the signal, the speed of information transmission and detection may be significantly less than the group speed. It is also known from information theory that only two points of the modulated sinusoid signal are required to determine its frequency, amplitude and phase. If the signal noise is small, these points can be very close together (nT $\leq$ t) and a sinusoidal curvefit can be performed to detect the signal. If information detection is based on this method, the speed of information detection may be only slightly less than the group speed. Note that applying this effect to the electric dipole will not eliminate the infinities in the phase and group speed curves; it will only reduce the width of the superluminal regions.

### *4.4  Magnetic dipole and oscillating gravitational mass*

Two other physical systems are noted to generate similar superluminal waves. Mathematical analysis of a magnetic dipole and a gravitationally radiating oscillating mass [3, 4, 5] reveals that they are governed by the same partial differential equation as the electric dipole. For the magnetic dipole, the only difference is that electric and magnetic fields are reversed. Consequently all of the analysis presented in this paper also applies to this system, and therefore similar superluminal wave propagation near the source is also predicted from theory.

For a vibrating gravitational mass, the difference is that electric (E) and magnetic (B) fields are replaced by analogs: the electric (G) and magnetic (P) component of the gravitational field [14]. In addition, a second mass vibrating with opposite phase is required to conserve momentum thereby making the source a quadrapole. But very close to the source, the effect of the second mass is negligible and can be neglected in the analysis. Consequently superluminal wave propagation is also predicted next to the source. Further away from the source the fields tend to cancel. Evidence of infinite gravitational phase speed at zero frequency has been observed by a few researchers by noting the high stability of the earth's orbit about the sun [15, 16]. Light from the sun is not observed to be collinear with the sun's gravitational force. Astronomical studies indicate that the earth's acceleration is toward the gravitational center of the sun even though it is moving around the sun, whereas light from the sun is observed to be aberated. If the gravitational force between the sun and the earth were aberated then gravitational forces tangential to the earth's orbit would result, causing the earth to spiral away from the sun, due to conservation of angular momentum. Current astronomical observations estimate the phase speed of gravity to be greater than $2 \times 10^{10}c$. Arguments against the superluminal interpretation have appeared in the literature [9, 10]

## 5  Conclusion

A simple experiment has been presented which shows that an oscillating electric dipole generates superluminal transverse electric field waves (phase and group) about one quarter wavelength outside the dipole source and that the waves travel superluminally toward and away from the source. The results have been shown to be consistent with electromagnetic theory. Arguments against this superluminal interpretation have been reviewed and shown to be deficient. Relativistic analysis indicates that from a moving observer's perspective, the superluminal signals generated by a stationary electric dipole can be seen to travel backward in time. Due to the mathematical similarity, two other physical systems are noted to have similar superluminal results: radiating magnetic dipole and oscillating gravitational mass.